\documentclass[12pt]{article}
\usepackage{amsmath, amssymb}
\usepackage{graphicx,psfrag,epsf}
\usepackage{enumerate}
\usepackage{natbib}
\usepackage{url} % not crucial - just used below for the URL 

\newcommand{\blind}{0}

\begin{document}

\def\spacingset#1{\renewcommand{\baselinestretch}%
{#1}\small\normalsize} \spacingset{1}

%%%%%%%%%%%%%%%%%%%%%%%%%%%%%%%%%%%%%%%%%%%%%%%%%%%%%%%%%%%%%%%%%%%%%%%%%%%%%%

\if0\blind
{
  \title{\bf Risk management in the use of published statistical results for policy decisions}
  \author{Duncan Ermini Leaf\hspace{.2cm}\\
    Leonard D. Schaeffer Center for Health Policy and Economics\\ University of Southern California}
\date{August 19, 2024}
  \maketitle
} \fi

\if1\blind
{
  \bigskip
  \bigskip
  \bigskip
  \begin{center}
    {\LARGE\bf Risk management in the use of published statistical results for policy decisions}
\end{center}
  \medskip
} \fi

\bigskip
\begin{abstract}
Statistical inferential results generally come with a measure of reliability for decision-making purposes.  For a policy implementer, the value of implementing published policy research depends critically upon this reliability.  For a policy researcher, the value of policy implementation may depend weakly or not at all upon the policy's outcome.  Some researchers might benefit from overstating the  reliability of statistical results.
Implementers may find it difficult or impossible to determine whether  researchers are overstating reliability.
This information asymmetry between researchers and implementers
can lead to an adverse selection problem where, at best, the full benefits of a policy are not realized or, at worst, a policy is deemed too risky to implement at any scale.  Researchers can remedy this by guaranteeing the policy outcome.  Researchers can overcome their own risk aversion and wealth constraints by exchanging risks with other researchers or offering only partial insurance.  The problem and remedy are illustrated using a confidence interval for the success probability of a binomial policy outcome.
\end{abstract}

\noindent%
\vfill

\newpage
\spacingset{1.45} % DON'T change the spacing!
\section{Introduction}
\label{sec:intro}

When statistical practice affects decisions, there is an interaction between a producer of statistical results and a consumer who makes a decision.  This paper focuses on a consumer's perspective in the policy statistics context where the producer is a policy researcher who publishes an estimate of a policy's effect and the consumer implements the policy in some population. 

Statistical results typically take the form of an estimate and a measure of uncertainty about the estimate.  The measure of uncertainty either directly or indirectly gives a measure of how reliable the estimate is for making policy decisions. Frequentist methods have $p$-values and confidence levels that imply controlled false-positive probabilities; 
Bayesian models have posterior probability for the effect of a policy or posterior predictive probability for the outcome of a policy in a population;
supervised machine learning  has various measures of out-of-sample error; binary prediction models have estimates of sensitivity, specificity, positive predictive value, and so on.

A policy implementer faces some cost for implementing a policy and does not realize the benefit of the policy until after those costs are paid.  Therefore, the implementer needs a reliable estimate of the policy's effect before making a decision to implement. In this way, the implementer consumes the statistical result produced by a researcher.

Researchers benefit from producing statistical results.  However, researchers might find it advantageous to misstate the reliability of their estimates when their benefits from producing estimates are not tightly bound to the estimates' actual reliability.  An implementation decision is more challenging when the implementer doubts a researcher's reliability claim.

Because statisticians sit on the production side of this relationship, it is natural for statisticians to think about improving reliability within the processes that generate statistical results.  This includes efforts to develop new methods and efforts to educate about better use of existing methods.  The effects of such efforts can be limited if they require both researchers and implementers to opt in.  Implementers might not have resources to invest in statistical methodology education.  Researchers also might not have those resources and those who do might not see a potential benefit in making the investment.

Before thinking about improving reliability on the production side, statisticians might find it more fruitful to consider consumers' perceptions of statistical results as a prerequisite problem.  This paper illustrates a way to help consumers distinguish reliable research from unreliable by placing some implementation risk onto researchers instead of solely upon implementers.  Statistical methodology improvements on the production side should be more highly valued under those conditions.

The consumer's fundamental problem is understanding a researcher's objective in reporting statistical results.  At first glance, the decision-theoretic model of statistical practice would seem to provide the answer:  statistical results limit the expectation of a loss function such as mean squared error or 0/1 loss.
This model has limited value to consumers for several reasons.  First,  consumers experience their own losses that are not necessarily reflected in the loss function. In fact, the loss function in published results can be selected without any input from consumers.  Second, the actual and potential losses experienced by a researcher are unknown to the consumer (and might even be unknown to the researcher).  As an example, ordinary least squares (OLS) estimates minimize the sum of squared residuals loss function.  However, suppose a researcher obtains an initial set of OLS estimates, drops some variables from the model to achieve an idea of parsimony, and then reports a final set of OLS estimates from this parsimonious model.  This researcher's true loss function is an interaction between the sum of squared residuals and unobservable thoughts about parsimony in the researcher's mind.\footnote{Even if the researcher provides a document describing their thought process, the consumer cannot be sure of its accuracy nor how observing the initial estimates influenced the parsimony criteria.}

In order to overcome decision theory's limitations, the discussion in this paper is based on a game-theoretic model of the interaction between a researcher and a policy implementer. 
While knowledge of game theory is not required, the basic set up is given here for readers who are familiar.
It is a sequential game.  First, Nature selects a value for the effect of a policy. This value is unknown to the researcher and implementer. The researcher plays next by publishing an estimate of the policy effect that maximizes the expectation of the researcher's utility function (negative loss function; denoted by $v$).  The utility function depends on the researcher's type.  Therefore, the researcher's choice of how to produce the estimate and, consequently, the reliability of the estimate depend on their type.  The researcher's type is unknown to the implementer, but is randomly selected from a distribution of types. The implementer may or may not have some estimate of this distribution.  The implementer plays last by choosing which scale of implementation is optimal considering the policy's uncertain outcome and uncertain reliability of the effect estimate.

Section \ref{sec:implementer_decision} of this paper gives a generic description of the decision problem faced by an implementer.  Section \ref{sec:perf_guarantee} introduces the policy performance guarantee as a way to overcome an implementer's uncertainty about the reliability of results. \citet{smaldino_natural_2016} argue that such guarantees could be ``highly punitive to individual researchers'' (p. 11),  which is undesirable because ``even high-quality research will occasionally fail to replicate'' (p. 12).
Section \ref{sec:perf_guarantee} explains how researchers can manage this risk for high-quality statistical results of known reliability.
Section \ref{sec:illustration} illustrates these ideas with a decision model for a policy with a binomial outcome. 
Section \ref{sec:discussion} discusses some limitations, implications, and extensions of these ideas.

\section{The implementer's decision problem}
\label{sec:implementer_decision}

Consider a policy implementer deciding whether or not to implement a policy based upon a researcher's published estimate of the policy's effect. The implementer must also decide the scale at which it should be implemented. In order to make these decisions, the implementer must take into account the estimated effect, the costs and potential benefits of the intervention,  uncertainty in the implementation, and uncertainty about the statistical results.  

The cost to implement the policy includes direct costs as well as the opportunity cost of giving up any current policies that would be replaced by the new policy. These costs are known (or estimated) when the implementer is making the implementation decision. However, before implementation, the benefit of the policy is known (or estimated) only as a function of the unknown policy outcome.  The actual benefit of the policy is not realized until after the policy is implemented and the outcome is observed.  Thus, from the implementer's perspective before implementation, the net benefit of the implemented policy is a random variable with a probability distribution defined by the unknown policy effect (the researcher's estimand) and uncertainty in the implementation. The policy is a failure if this random variable's realized value is negative. 

Setting aside uncertainty about the policy effect estimate, the implementer faces two sources of uncertainty in the implementation.  First, the implementation must match the policy and population described by the published estimate 
in order to achieve the same underlying effect.  
The net benefit of the policy depends on how a deviation from this target affects the policy outcome distribution.
Second, the implementer expects random variation in individual outcomes even if the underlying policy effect is known with certainty and the policy is correctly implemented.

The implementer must also contend with uncertainty about the policy effect estimate. 
The implementer expects highly reliable estimates to be close to the true effect of the policy in the study population, but the implementer knows that even highly reliable estimates still have some amount of uncertainty. The best the implementer can hope for is a high probability that the estimate is close to the true effect.  Then, in the long run, most implemented policies will be successful.

Putting all of this together, the implementer wants a decision rule that implements the policy when the researcher's estimate is probably close to a true effect that would probably yield a favorable policy outcome.  The decision rule 
has a random net benefit before implementation: zero net benefit if it chooses to not implement the policy and the random net benefit from the policy outcome if the rule chooses implementation.
The choice between these two is a function of the  researcher's estimate, which is a also a random variable.    
The implementer selects a decision rule that minimizes risk based upon some characteristic of the overall distribution of the decision rule's outcome, combining both implementation uncertainty and uncertainty of the estimate.  In what follows, the implementer considers the expected value of the decision rule.

In order to develop this decision rule, the implementer needs some information about the distribution of the researcher's estimate.  The researcher's measure of reliability summarizes this distribution.  As an example, suppose a researcher  reports a $95\%$ confidence interval for the policy's effect within which the implementer sees a positive expected net benefit. If the implementer believes the researcher reported their confidence level truthfully, the implementer could conclude that there is at most a $.05$ probability of observing the reported positive benefit when the policy's expected net benefit is actually negative. The implementer can use confidence interval inclusion as the decision rule and use the confidence level to assess the risk from using the decision rule. 

Unfortunately, the implementer does not know the actual reliability.  The only information immediately available to the implementer is the researcher's \emph{reported} reliability.  
Suppose the implementer suspects the coverage probability of the reported $95\%$ confidence interval is  actually less than $.95$ and, consequently, the probability of a false positive is greater than $.05.$  If the false positive probability is too large, the expected value of the decision rule could be negative. In that case, the implementer would not go forward with implementation at all or would reduce risk by scaling back the implementation to a smaller population.  The next section explains why the implementer could be doubtful of the researcher's reported reliability.

\subsection{The researcher population}
\label{sec:researcher}
The researcher who publishes a policy effect estimate is a member of a population of researchers.
Variation in the reliability of researchers' estimates depends on group- and individual-level characteristics of the researcher population.
At the group level, reliability can decrease as the number of researchers studying the same policy increases \citep{ioannidis_why_2005}.  Individual-level variation comes from the costs and benefits each researcher faces and how different researchers value them.  Generally, there is an opportunity cost for individuals to engage in a research project instead of other research projects or non-research activities.  Then, within a research project, there are several sources of costs and benefits of reliability.

It can be costly for researchers to report the reliability of statistical results truthfully due to the availability of relatively convenient approximations.  For example, asymptotic theory provides widely used bounds for confidence intervals with approximately $95\%$ coverage probability even though the minimum coverage probability might be substantially less than $95\%$ (for example, see \citeauthor{brown_interval_2001}, \citeyear{brown_interval_2001}).  The effort required to determine the actual minimum coverage probability is a cost to the researcher. 

More generally, much of statistical theory is based upon assuming probability models for an unobservable data generation process.  Reliability measures derived from misspecified models will be inaccurate. However, more robust or less biased methods can be costly to researchers in terms of additional education needed to understand the methods, additional computational requirements, measurement of additional variables, and potentially larger sample size requirements. Low-cost options include assuming $\log$-Normality of positively-valued data, treating a complex survey as a simple random sample, dropping missing data cases, and ignoring confounding.

The cost of truthful reporting also comes from quality control processes. Errors in data coding or computer programming affect the reliability of results.  Quality control processes that check for these errors take a researcher's time and resources away from other activities that could be more beneficial to the researcher.

While truthfully reporting reliability has a cost, the {\it perception} of reliability can be beneficial to a researcher.
In commercial settings, company valuations may depend on the perceived reliability of the research underlying a product.
Among academic researchers, career advancement  and financial conflicts of interest can lead to overstated reliability \citep{ioannidis_why_2005,smaldino_natural_2016}. 
Therefore, a researcher's cost for truthfully reporting reliability may also include foregone financial benefits or other career advancement opportunities.

Researchers face a potential cost if their research is implemented as policy and it fails or a potential benefit if it is a success.  This varies depending on the researcher's specific circumstances and their attitude toward the future. One researcher might feel no effect from a policy failure where another would find it very costly.

The researchers' costs and benefits discussed so far depend on external conditions that are observable to some extent.  A researcher can also derive unobservable emotional costs and benefits from how they conduct research.  A researcher might gain a sense of well-being knowing that their work is of high quality independent of any implementation outcome.  A researcher might experience fear of mistakes or a sense of shame around potential policy failures even when there would be no material cost to the researcher.
This is internal to the researcher and varying across researchers.  Thus, two researchers facing the same external costs and benefits, valuing those costs and benefits equally, might report reliability differently.

\subsection{Information asymmetry between researcher and implementer}
\label{sec:info_asym}
In order to make an implementation decision, ideally, the implementer would make their own determination of reliability by answering three questions about the published statistical results:
%\begin{enumerate}
%	\item[\textbf{Q1}:] Are the models and methods appropriate for the data generating process studied by the researcher?
%	\item[\textbf{Q2}:] Were the methods applied correctly?
%	\item[\textbf{Q3}:] How do the statistical results generalize to the implementation population?
%\end{enumerate}
	(Q1) Are the models and methods appropriate for the data generating process studied by the researcher?
	(Q2) Were the methods applied correctly?
	(Q3) How do the statistical results generalize to the implementation population?
Answering Q1 requires access to the data, knowledge of the data collection process, and expertise in both the subject matter and statistical methods.  Q2 requires access to the data and knowledge of the analytical steps used to produce the statistical results.  This includes understanding computer code in whichever language it was written.
Q1 and Q2 are prerequisites for answering Q3. Q3 additionally requires knowledge of statistical theory and knowledge of the implementation population.

The implementer faces a cost in attempting to answer Q1--Q3.  For example, Q1 and Q3 might require the implementer to obtain a Ph.D.-level understanding of statistics or machine learning while Q2 might require learning a new programming language.\footnote{\emph{Implementer} loosely refers to an individual or an organization.  Someone within an implementing organization would have to obtain Ph.D.-level understanding or learn a programming language.  Alternatively, the organization would have to hire someone with this knowledge.}
Answering Q2 might also require paid access and training in computing environments like a secure data enclave (for example, Centers for Medicare and Medicaid Services' Virtual Research Data Center) or high-performance computing (for example, Amazon Elastic Compute Cloud).

Some information is unavailable to the implementer at any cost.  In answering Q1, even if a researcher shares their data, the implementer cannot observe the already-completed data collection process.  For example, randomization failures in an experiment are not directly observable in the data. Data manipulation is also unobservable: an implementer cannot determine from the data alone whether some cases were modified, omitted, or imputed.

When the implementer is limited in their ability to learn the reliability of a researcher's statistical results, the implementer could consider the reliability as a realization from the distribution of reliability in the population of researchers.  Learning this distribution requires the implementer to find an unbiased and sufficiently large sample of replicated results. If none exists, the implementer must perform the replication studies.  As such, the implementer's cost to learn the distribution may far exceed the cost to directly investigate the reliability of the published results.  The cost to learn this distribution may even exceed the cost to implement the policy itself.

Having a limited ability to learn the reliability of a published result or the distribution from which it comes, the implementer could turn to third-party estimates of reliability. See \citet*{altmejd_predicting_2019} and \citet*{gordon_are_2020} for examples.  However, because these reliability estimates are research products like the initial policy research under consideration, the implementer must deal with the uncertain reliability of a reliability estimate. The implementer faces an infinite regress in using third-party estimates of reliability.

In the end, an implementer may be left with a poorly estimated probability distribution for reliability.  
This distribution combines their own and others' previous experience, subject matter knowledge, and published estimates of reliability (also of uncertain reliability).  
In a Bayesian sense, this is the implementer's \emph{prior distribution} over the researcher's reliability before observing the outcome of the policy in question.
The cautious implementer prefers that their distribution underestimate reliability rather than overestimate it.
In the safest extreme, the implementer would make the implementation decision under the assumption that the policy will certainly be a total failure.  
However the distribution is formed, the implementer has no better alternative for determining the risk of implementing the policy.

\subsection{Adverse selection}
The relationship between a researcher and an implementer is analogous to the relationship between the seller and potential buyer of a used car.  The buyer wants to know whether the car has any major mechanical problems that will be expensive to repair.  If the seller says the car has no problems, there are three possibilities: the car truly has no problems, the car has problems that the seller is unaware of, or the seller is knowingly concealing the problems so that the buyer might pay more money for the car.  If the buyer cannot easily determine the car's mechanical condition, the buyer must consider the probability of potential problems and the cost of repair in deciding whether or not to purchase the car.  \citeauthor{akerlof_market_1970}'s (\citeyear{akerlof_market_1970}) analysis of the car market showed that the presence of untruthful sellers and the inability of buyers to determine quality can lead to adverse selection: even though there are sellers of high-quality cars and buyers willing to pay for them, the high-quality cars are not sold.

Now consider a researcher who truthfully reports reliability.  When they report a $95\%$ confidence interval, the minimum coverage probability is at least $.95$.  Suppose an implementer would find coverage probability of at least $.9$ acceptable.  When the implementer cannot determine that the researcher is truthful, the implementer turns to the probability distribution described in Section \ref{sec:info_asym}.  From this, the implementer might conclude that the average coverage probability in this population of researchers is $.8$ --- an unacceptably low level.  Although the researcher has $.95$ coverage probability and the implementer would be willing to implement the policy based upon this interval with $.95$ coverage probability, the policy is not implemented.

\subsection{A remedy: performance guarantees}
\label{sec:perf_guarantee}
\citet{akerlof_market_1970} discusses several remedies for the adverse selection problem in the car market.  One remedy is a performance guarantee offered by the seller.  This paper proposes the performance guarantee as a remedy for adverse selection in policy research implementation. Here, a performance guarantee acts as a product warranty or an insurance policy for the outcome of the implemented policy.

From the implementer's perspective, there are two kinds of researchers who are never willing to guarantee the policy outcome: researchers who have greatly overstated the reliability of their results and researchers who truthfully reported reliability, but do not think their estimate is applicable to the implemented policy.  The implementer would like to avoid these cases.

There is an additional type of researcher who might be initially unwilling to offer a guarantee.  This researcher produced reliable results that generalize to the implementation population.  Yet, the risk of policy failure is still unacceptable in relation to the researcher's existing wealth or their attitude toward risk. There are three potential ways the researcher can work around this in order to offer the guarantee.  First, the researcher can pay a third party to take some of the risk --- paying a certain amount today to eliminate an uncertain future outcome. Second, the researcher could find another researcher who needs to guarantee some other policy outcome. These two could exchange a fraction of their risks.  By diversifying the risks, each researcher faces a lower probability of extreme outcomes. Third, the researcher can offer insurance against only the most extreme policy failures or insurance against only a proportion of the risk.  These  strategies are illustrated in Section \ref{sec:researcher_risk_mgmt} and extensions are discussed in Section \ref{sec:discussion}.

\section{Illustration with a binomial policy outcome}
\label{sec:illustration}
Section \ref{sec:implementer_decision} described the implementation decision problem in general terms.  This section illustrates those ideas using a decision model for a binomial policy outcome.  In this model, the implementer encounters a single published study in which the study population matches  the implementation population. Section \ref{sec:discussion} discusses the case of multiple studies on the same population or a mismatch between the study population and the implementation population.  Section \ref{sec:ill_no_guarantee} shows the decision problem faced by the implementer when the researcher makes no guarantee of the policy outcome.  Section \ref{sec:ill_adverse_selection} provides examples of how adverse selection might arise in this model.  Section \ref{sec:ill_guarantee} illustrates how the implementer's decision changes when the researcher guarantees the policy outcome. Finally, Section \ref{sec:researcher_decision} demonstrates how researchers can manage their own risk when deciding whether or not to offer a guarantee in this model.

\subsection{The implementer's decision \emph{without} a guarantee}
\label{sec:ill_no_guarantee}
Consider a policy intervention on individuals with binary outcomes each having probability $p$ of success.  For example, the policy could be a vaccine outreach strategy in a population that is unlikely to be vaccinated without an intervention.  Success is an individual being vaccinated.  A researcher publishes a one-sided confidence interval for $p$ with lower bound $L$, claiming the post-intervention vaccination rate is at least $L$ with $(1-\alpha')\%$ confidence.

The implementer --- a public health agency in the vaccine outreach example --- must decide whether or not to implement the policy based upon the confidence interval reported by the researcher.  The implementer understands the definition of a $(1-\alpha)\%$ confidence interval implies: 
\begin{equation}
\label{eqn:ci_coverage}
\mathrm{Pr}(L \leq p) \geq 1-\alpha
\end{equation}
for all $p\in[0,1]$.  Beyond this, the implementer has no additional information about the unknown value of $p$ nor the sampling distribution of $L$.

The implementer's cost of implementing the policy for $m$ individuals is $c_m$.  
Assume members of the target population are independent so that the number of successes, $X_m,$ is a binomial random variable: $X_m\sim \mathrm{Binomial}(m,p)$. The benefit to the implementer of $x$ successes is a function $b(x)$.  Assume $b(x)$ is increasing in $x$ and $b(0) = 0.$
For the vaccine outreach policy, $b(x)$ could be the expected reduction in medical costs for $x$ vaccinated patients relative to $x$ unvaccinated patients.  Thus, the net benefit to the implementer is $b(X_m) - c_m$ with expected net benefit $\mathrm{E}[b(X_m)] - c_m.$

The implementer serves a population of $M$ individuals in which the policy can be implemented. This could be determined by the actual size of the population or by the implementer's budget constraints. 
To keep this illustration simple, suppose for all $p\in[0,1]$ there exists an $m^*(p) \in [1,M]$ such that $\mathrm{E}[b(X_m)] < c_m$ for $m < m^*(p)$ and $\mathrm{E}[b(X_m)] - c_m$ 
is strictly increasing in $m$
for $m \geq m^*(p)$. In other words, if there is a positive expected net benefit for some number of individuals, then the implementer maximizes the expected net benefit by intervening on as many individuals in the population as possible (subject to any budget constraints).

Let $p_0$ be the break-even success rate when the policy is applied to the entire population such that $\mathrm{E}[b(X_M)] = c_M$ when $X_M\sim\mathrm{Binomial}(M,p_0).$
Suppose the implementer's decision rule implements the policy only when the confidence interval bound satisfies $L > p_0.$  If $L \leq p_0$, the policy is not implemented and there is no cost to the implementer.  The net benefit of applying this decision rule to $m$ individuals is the random variable,
\[
U^0_m = \mathbb{I}\{L > p_0\}(b(X_m) - c_m),
\]
where $\mathbb{I}$ is the $0/1$ indicator function.  The expected net benefit of the decision rule is:
\begin{equation}
\mathrm{E}[U^0_m] = \mathrm{Pr}(L > p_0) \mathrm{E}[b(X_m) - c_m].
\label{eqn:drenb}
\end{equation}

The implementer cannot determine the expected net benefit in (\ref{eqn:drenb}) because they do not know $p$ nor the sampling distribution of $L.$  However, they can limit their expected loss by reasoning around the worst case as follows. Let $\alpha=\sup_{p < p_0} \mathrm{Pr}(L > p_0)$ be the maximum false positive probability of the decision rule (the probability that the policy is implemented when it has a negative net benefit). In the worst case ($p=0$) there will be no successes   and the implementer will face a certain loss of $c_M.$ Therefore,
\begin{equation}
\inf_{p < p_0} \mathrm{E}[U^0_m] \geq -c_m \sup_{p < p_0} \mathrm{Pr}(L > p_0) = -c_m \alpha, 
\label{eqn:worst_case_no_guarantee}
\end{equation}
gives a worst-case lower bound for the expected loss.  This might seem like an extreme overestimate, but it is the best the implementer can do  without additional information.

As a complementary cumulative distribution, $\mathrm{Pr}(L > p_0)$ is decreasing in $p_0.$ If the researcher reports their coverage probability $(1-\alpha')$ truthfully, then (\ref{eqn:ci_coverage}) implies that $\mathrm{Pr}(L > p_0) \leq \alpha'$ when $p < p_0$  and so $-c_m \alpha' $ is the lower bound for the expected loss.   If $L > p_0$ and the implementer has a minimum tolerable expected loss, $\underline{u} < 0,$ then the policy is applied to the largest number of individuals $m$ such that $-c_m \alpha' \geq \underline{u}$.  If $-c_1 \alpha' < \underline{u}$ the policy is not implemented at all.

All of this assumes the researcher reported their true coverage probability. However, for the reasons discussed in Section \ref{sec:implementer_decision}, the implementer does not know whether the true coverage probability matches the nominal level reported by the researcher.  If the researcher overstates their coverage probability, then $\alpha' < \alpha$ and the implementer does not know whether the bound for their expected loss, $-\alpha c_m,$ exceeds $\underline{u}.$  
If the implementer is unable to determine a value for $\alpha$, then they can use their expectations, if any, about the researcher population in order to make a decision on whether or not implement and at what scale. The next section gives some examples of researchers who overstate their coverage probability and how an implementer might respond.

\subsection{Adverse selection examples}
\label{sec:ill_adverse_selection}
The following examples illustrate how an implementer might reason probabilistically about the false positive probability, $\alpha.$  These examples also show how adverse selection might arise as a consequence.  Example 1 involves researchers who report fraudulent results in order to influence implementers.  Example 2 involves poor methodology that is convenient for some researchers, but not necessarily intended to manipulate implementers.  These toy examples serve only to illustrate the problems at hand. Reality is much more complex.

In both examples, the implementer does not understand what the researcher has done.  The implementer is given only a confidence statement: $L \leq p$ with $(1-\alpha')\%$ confidence.  These examples show the correct probability distribution for $\alpha'$ that the implementer should use for making an implementation decision.
As discussed in Section \ref{sec:info_asym}, many implementers would not be able to determine these distributions and would err on the side of caution. Consequently, the adverse selection would be worse than in these examples.  

\subsubsection{Example 1: fraudulent researchers}
Suppose untruthful researchers want to see the policy implemented and therefore want to report a lower bound $L > p_0.$ However, the value of $p_0$ is unknown to the researcher before they publish.  They start by guessing a value $p'_0$ for $p_0.$  Their guessing strategy gives $p'_0$ only .5 probability of exceeding $p_0.$ Next, they compute a confidence interval lower bound having at least $1-\alpha'$ coverage probability.  If their lower bound is greater than $p'_0,$ they report it.  If their lower bound does not exceed $p'_0$, they report $L = p'_0$ as the lower bound with an untruthful claim of $(1-\alpha')\%$ confidence.    This process means the interval from an untruthful researcher gives the decision rule a  false positive probability of $.5\alpha' + .5.$  If the implementer believes there is .25 probability that the interval came from a untruthful researcher, then the implementer faces a false positive probability of $\alpha = .875\alpha' + .125$ at nominal confidence level $(1-\alpha')\%.$  

The minimum achievable false positive probability for the implementer in this example is .125 as the nominal probability goes to zero.  Suppose the implementer's minimum acceptable loss is $\underline{u} = -.05 c_M.$ This means they would implement the policy if $L > p_0$ in a confidence interval with true coverage probability of at least $.95.$  Now suppose a truthful researcher publishes a $99\%$ confidence interval with $L > p_0.$  If the implementer could determine that the researcher is truthful, the policy would be implemented.  However, when they cannot distinguish truthful researchers from untruthful, the false positive probability of the decision rule is around $.134$ at the $99\%$ nominal level.  This gives a minimum expected loss of $-.134 c_M$ to implement the policy in the full population.  Since this exceeds $\underline{u}$, the implementer scales back the intervention population size to the largest $m$ such that $-.134 c_m \geq -.05 c_M$.  This amounts to a roughly two-thirds reduction in the cost of implementation.  If there is a constant implementation cost for each additional individual to whom the policy is applied ($c_m$ is linear in $m$), then at least two thirds of the eligible population do not experience potential benefits from the policy due to the presence of untruthful researchers.  If the implementation has a large fixed cost such that $c_1 / c_M > .374,$ then the policy is not implemented at all.

\subsubsection{Example 2: poor methodology and publication bias}
In this example, a researcher performs a randomized controlled trial to evaluate a new policy against the status quo policy.  Both the control and treatment groups have 300 participants.  Let $p_C$ be the success probability of the status quo policy in the control group and let $p$ be the success probability of  the new policy in the treatment group.  Correspondingly, let $\hat{p}_C$ and $\hat{p}$ be the success rates observed in each sample group.  Untruthful researchers first perform a two-sample test of $\mathrm{H}_0: p \leq p_C$ vs. $\mathrm{H}_1: p > p_C,$ rejecting $\mathrm{H}_0$ when
\[
\frac{\hat{p} - \hat{p}_C}{\sqrt{2 p_z (1-p_z)/300}} \geq \Phi^{-1}(1-\alpha'),
\]
where $p_z = (\hat{p}_C + \hat{p})/2$ and $\Phi$ is the standard Normal cumulative distribution function.  When $\mathrm{H}_0$ is rejected, untruthful researchers publish a $(1-\alpha')\%$ confidence bound for $p$ using the Wald approximation:
\[
L = \hat{p} - \Phi^{-1}(1-\alpha')\sqrt{\hat{p} (1-\hat{p})/300}.
\]
An untruthful researcher publishes the result only when they reject $\mathrm{H}_0$.  Note that the untruthful researchers in this example are not actively trying to deceive the implementer.  

\begin{figure}
\begin{center}
\includegraphics[width=4.25in]{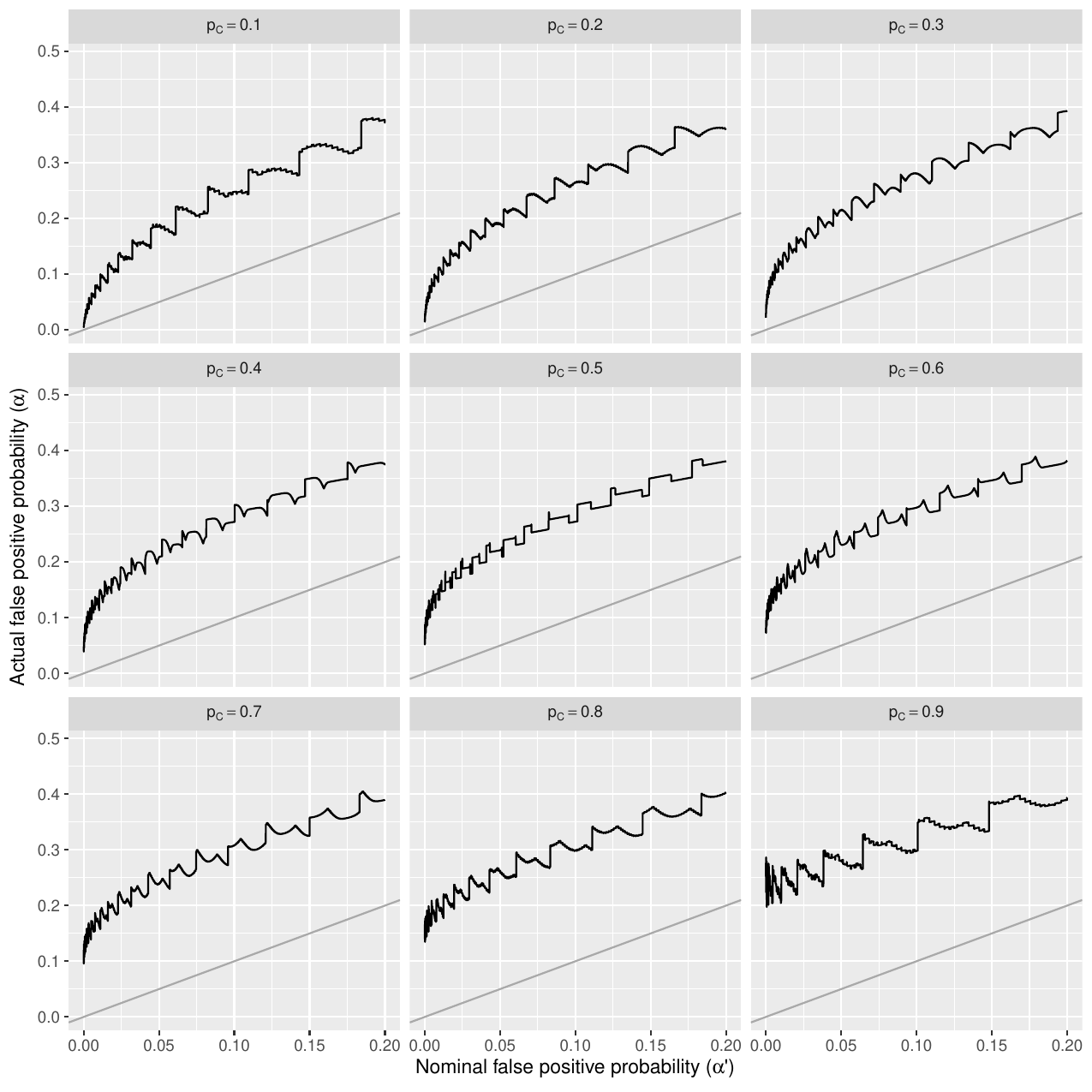}
\end{center}
\caption{Researcher's nominal false positive probability ($\alpha'$) and implementer's actual false positive probability ($\alpha = \sup_{p < p_0} \mathrm{Pr}(L > p_0)$) for different values of $p_C$ in Example 2.  The dotted line represents truthful reporting by all researchers ($\alpha' = \alpha$). \label{fig:ex2fpr}}
\end{figure}

Assume the implementation population is already experiencing the status quo policy. For simplicity, assume the only cost of implementing the new policy is the opportunity cost of giving up the status quo policy.  Thus, the implementer wants to switch to the new policy when $p > p_C.$  Suppose the implementer believes there is probability .5 that the interval came from an untruthful researcher.  Figure \ref{fig:ex2fpr} shows the false positive probability when the new policy is inferior to the status quo policy ($p < p_C$) for different values of $p_C.$

Now suppose the status quo policy has $p_C = .5$ as the probability of success. As in the previous example, consider an implementer who has a minimum acceptable loss of $\underline{u} = -.05 c_M,$ meaning a truthful $95\%$ confidence interval with $L > .5$ is required for implementation.  However, as shown in the center panel of Figure \ref{fig:ex2fpr}, a nominal $.05$ false positive probability corresponds to an actual $.22$ false positive probability.  In order for the implementer to achieve an actual false positive probability of at most $.05$, they would need the published interval to have at least $97.5\%$ nominal confidence level.

Suppose the researcher's published interval has only $95\%$ nominal coverage. The policy is not implemented for the full population because the minimum expected loss of $-.22 c_M$ exceeds the implementer's limit.  The implementer scales down the implementation to $m$ individuals such that $c_m \leq .227 c_M$ and does not implement the policy at all if $c_1 > .227 c_M.$  If the implementation cost is linear in $m,$ then at least three quarters of the population do not experience potential benefits due to the presence of untruthful researchers. 

\subsection{The implementer's decision \emph{with} a guarantee}
\label{sec:ill_guarantee}
The fundamental problem in the previous examples is that the implementer cannot distinguish truthful researchers from untruthful researchers.  This section shows how researchers who guarantee the policy outcome ameliorate this problem for the implementer. 

Let $Y_m = b(X_m) - c_m$ be the net benefit of the policy to the implementer. Let $Y_m^+ = Y_m\mathbb{I}\{Y_m > 0\}$ and  $Y_m^- = Y_m\mathbb{I}\{Y_m < 0\}.$ A researcher can offer a performance guarantee when $L > p_0$  by agreeing to pay the implementer $-Y_m^-$ if the policy fails.  With this guarantee, the implementer's net benefit from the decision rule becomes the random variable,
\[
U_m^1 = \mathbb{I}(L > p_0)(Y_m - Y_m^-) = \mathbb{I}(L > p_0)Y_m^+.
\]
Since $\mathrm{E}[U_m^1]$ is non-negative for any $p$, the policy is implemented whenever $L > p_0$. 

The researcher's offer to pay $-Y_m^-$ is full insurance against a policy failure.  More generally, a researcher could offer only partial insurance against a loss exceeding some amount $k \in (-c_m,0)$ of the total loss. If the researcher pays the implementer $-(Y_m-k)^-$ when the policy fails, the worst-case  bound for the implementer's  expected loss becomes
\[
\inf_{p < p_0} \mathrm{E}[\mathbb{I}(L > p_0)(Y_m - (Y_m-k)^-)] \geq k \sup_{p < p_0} \mathrm{Pr}(L > p_0). 
\]
If the implementer has no information about $\mathrm{Pr}(L > p_0),$  they would take $\sup_{p < p_0} \mathrm{Pr}(L > p_0) =1$ and the worst-case expected loss is simply $k.$  The implementer will still use the decision rule if $k \geq \underline{u}.$  
The implementer's decision is more challenging when $k < \underline{u}$ because the implementer must consider the false positive probability in the population of researchers who would insure losses beyond $k.$  Consequently, the implementer might scale back or not implement at all when $k < \underline{u}.$ This is discussed further in Section \ref{sec:partial_insurance}.

Another form of partial insurance is for the researcher to reimburse only a proportion $s \in (0,1)$ of the total loss, leaving the implementer to pay the remaining $(1-s)Y_m^-$. The  $s = 1-\alpha'$ case is particularly interesting.  In this case, the researcher pays $-(1-\alpha')Y_m^-$ to the implementer, giving the implementer an expected net benefit of 
\[
\mathrm{E}[\mathbb{I}(L > p_0)(Y_m + (1-\alpha')Y_m^-)] = \mathrm{Pr}(L > p_0)\mathrm{E}[Y_m^+ -\alpha'Y_m^-].
\]
At worst, this expected net benefit is bounded as
\[
\inf_{p < p_0} \mathrm{Pr}(L > p_0)\mathrm{E}[(Y_m^+ -\alpha'Y_m^-)] \geq - c_m \alpha' \sup_{p < p_0}\mathrm{Pr}(L > p_0) \geq -c_m \alpha' .
\]
From the implementer's perspective, this level of insurance is at least as good as the researcher truthfully reporting their false positive probability in (\ref{eqn:worst_case_no_guarantee}).

\subsection{The researcher's decision to offer a guarantee}
\label{sec:researcher_decision}

As discussed in Section \ref{sec:researcher}, the researcher faces many potential costs and benefits in reporting their coverage probability truthfully.  Let $V_m^0$ be the net benefit to the researcher from publishing their statistical results and having the policy implemented in a population of size $m$ individuals. The net benefit of publishing, but never seeing the policy implemented is $V_0^0$.  In a commercial setting where the implementer pays the researcher directly for implementation, $V_m^0$ for $m > 0$ includes a transfer of $c_m$ from the implementer.  In an academic setting, $V_m^0$ may include the value of career advancement, potential salary increases, expanded access to research funding, new consulting opportunities, and returns from commercialization. 
In both settings, $V_m^0$ includes direct and opportunity costs of engaging in research.
$V_m^0$ also includes any negative consequences of policy failure.  Since these are not realized before the time of implementation, $V_m^0$ is a random variable.  The implementer does not know the probability distribution for each $V_m^0$ nor how exactly the researcher makes decisions using these distributions.  However, the researcher must prefer receiving $V_m^0$ over receiving nothing for some $m \geq 0$ or else the researcher would not have published in the first place.  For all $m,$ assume the distribution of $V_m^0$ depends on the interval bound, $L,$ and the reported confidence level, $(1-\alpha')\%.$

Let $V_m^1$ be the net benefit to the researcher who offers to fully guarantee the policy outcome.  When the policy is implemented $(m \geq 1),$ then $V_m^1 = V_m^0 + Y_m^-.$  If policy is not implemented, then the net benefit is $V_0^1 = V_0^0$ --- the same as when no guarantee is offered.  If the implementer will not implement without a full guarantee, the researcher faces a choice between not offering the guarantee and receiving $V_0^1$ or offering the guarantee and receiving $V_m^1$ for the implementers choice of $m$ when $L > p_0.$ 

Suppose the researcher is risk-averse.\footnote{
	Assuming risk aversion is reasonable here because risk seekers seem unlikely to work on research that is weakly tied to outcomes.  Also, risk seekers who produce reliable results have less need for the risk-management strategies discussed in this section.}  
They evaluate random outcomes according to expectations over an increasing, concave utility function $v$ \citep{pratt_risk_1964}.  
For a given value of $p,$ the researcher who reports $L > p_0$ has expected net benefit $\mathrm{E}[v(V_m^1) | L > p_0]$ if they offer a guarantee and $\mathrm{E}[v(V_0^1) | L > p_0]$ if they do not.  
This gives a necessary condition that that $\mathrm{E}[v(V_m^1) | L > p_0] \geq \mathrm{E}[v(V_0^1) | L > p_0]$ for some $p$ in order to offer a guarantee.
Now suppose the researcher has a lower limit, $\underline{v},$ for their expected value from participating in research.  This gives a second necessary condition for the researcher to offer a guarantee: 
\begin{equation}
\label{eqn:participation2}
\mathrm{Pr}(L \leq p_0)\mathrm{E}[v(V_0^1) | L \leq p_0] + \mathrm{Pr}(L > p_0)\mathrm{E}[v(V_m^1) | L > p_0] \geq \underline{v}
\end{equation}
for all $p.$  These necessary conditions give some insight into the types of researchers who offer guarantees.

If both $\mathrm{E}[v(V_0^1) | L \leq p_0] \geq \underline{v}$ and $\mathrm{E}[v(V_m^1) | L > p_0] \geq \underline{v}$ for all $p,$ then (\ref{eqn:participation2}) is satisfied.  This means that $\mathrm{E}[v(V_m^0 - Y_m^-) | L > p_0] \geq \underline{v}$ for all $p.$  Since $\mathrm{E}[v(V_m^0 - Y_m^-) | L > p_0] \leq v(\mathrm{E}[V_m^0 | L > p_0]  - \mathrm{E}[Y_m^-])$ by the concavity of $v$ and independence of $L$ and $Y,$ then $\mathrm{E}[V_m^0 | L > p_0]  - \mathrm{E}[Y_m^-] \geq v^{-1}(\underline{v})$ must hold, even in the worst case for $Y^-$ when $p \approx 0.$  In other words,
the magnitude of the researcher's minimum acceptable expected loss, $|v^{-1}(\underline{v})|,$ must far exceed the cost of implementation or the researcher's expected net benefit from extreme policy failures must be large.  While these cases seem unlikely to occur in practice, they cannot be controlled by the implementer when they do occur.  However, the implementer is indifferent if these researchers fully insure the policy outcome.

When either $\mathrm{E}[v(V_0^1) | L \leq p_0] < \underline{v}$ or $\mathrm{E}[v(V_m^1) | L > p_0] < \underline{v}$ for some $p,$ then (\ref{eqn:participation2}) gives two necessary conditions on  $\mathrm{Pr}(L > p_0)$ in order for the researcher to offer a guarantee.  First, for $p$ such that $\mathrm{E}[v(V_m^1) | L > p_0] < \underline{v} \leq \mathrm{E}[v(V_0^1) | L \leq p_0],$ then
\begin{equation}
\label{eqn:fp_control}
\mathrm{Pr}(L > p_0) \leq \frac{\mathrm{E}[v(V_0^1) | L \leq p_0] - \underline{v}}{\mathrm{E}[v(V_0^1) | L \leq p_0] - \mathrm{E}[v(V_m^1) | L > p_0]}
\end{equation}
must hold.
Second, for $p$ such that $\mathrm{E}[v(V_0^1) | L \leq p_0] < \underline{v} \leq  \mathrm{E}[v(V_m^1) | L > p_0],$ then
\begin{equation}
\label{eqn:tp_control}
\mathrm{Pr}(L > p_0) \geq \frac{\mathrm{E}[v(V_0^1) | L \leq p_0] - \underline{v}}{\mathrm{E}[v(V_0^1) | L \leq p_0] - \mathrm{E}[v(V_m^1) | L > p_0]}
\end{equation}
must hold.  Across $p$ and $p_0,$ these two conditions limit the acceptable sampling distributions for $L$ among researchers who offer a guarantee. 

The first condition above (\ref{eqn:fp_control}) puts a limit on the researcher's false positive probability when $\mathrm{E}[v(V_m^1) | L > p_0]$  is increasing in $p.$  This serves to filter out unreliable results, though the implementer is indifferent when the policy outcome is fully insured by the researcher.  The second condition (\ref{eqn:tp_control}) requires the statistical method to have enough power in order for the researcher to offer a guarantee.  This is bad for both parties when it precludes implementation of mutually beneficial research from researchers who are highly risk-averse (represented by $v$) or have a very tight budget with only a small expected benefit from implementation ($\underline{v} \approx \sup_{p > p_0}\mathrm{E}[v(V_m^1) | L > p_0]$).  However, some researchers can get around this by employing their own risk management strategy.

\subsubsection{Risk management strategies for the researcher}
\label{sec:researcher_risk_mgmt}
The general problem faced by some researchers is that $\mathrm{E}[v(V_m^1) | L > p_0]$ is not large enough even when $p > p_0.$  Consequently, they do not offer a guarantee and the beneficial policy is not implemented.  Risk management for the researcher amounts to replacing $V_m^1$ with an alternative $W$ such that $\mathrm{E}[v(W) | L > p_0]$ is large enough to justify the guarantee.

One approach 
involves transferring risk to a third party.  This transfer could be a purchase of insurance.  For example, if the researcher can pay $t$ to transfer a proportion $(1-\gamma)$ of $Y_m^-$ to the third party, then  $\mathrm{E}[v(V_m^1) | L > p_0]$ is replaced with $\mathrm{E}[v(V_m^0 + \gamma Y^-_m - t) | L > p_0].$ The risk transfer could also be an exchange. Suppose the third party is another researcher who has risked $Z^-$ to guarantee their own results for some other policy.  If a proportion $(1-\gamma_1)$ of $Y_m^-$ is exchanged for a proportion $\gamma_2$ of $Z^-,$ then the original researcher has replaced $\mathrm{E}[v(V_m^1) | L > p_0]$ with $\mathrm{E}[v(V_m^0 + \gamma_1 Y^-_m + \gamma_2 Z^-) | L > p_0].$  

As an alternative to transacting with a third party, the researcher can also manage risk by offering the implementer only partial insurance. As mentioned in Section \ref{sec:ill_guarantee}, the researcher can insure tail losses beyond some amount $k<0$ or insure only a proportion $s < 1$ of any losses.  These strategies replace  $\mathrm{E}[v(V_m^1) | L > p_0]$ with $\mathrm{E}[v(V_m^0 +  (Y_m-k)^- | L > p_0]$ and $\mathrm{E}[v(V_m^0 +  s Y_m^-) | L > p_0],$ respectively.
The researcher minimizes their risk while being certain of implementation in $m$ individuals by choosing $k = \underline{u}$ or $s = -\underline{u}/c_m.$

\section{Discussion, limitations, and extensions}
\label{sec:discussion}

The model and decision processes presented so far are limited by assuming: (i) the study population and implementation population are the same; (ii) the distribution of the policy outcome ($X_m$) is known to the researcher and implementer; (iii) the implementer knows the cost and benefits of the policy with certainty; and (iv) there is only a single published estimate of the policy's effect. The remainder of this section discusses these limitations and how to move beyond them. Additionally, Section \ref{sec:researcher_adverse} discusses the full extent of the adverse selection problem and Section \ref{sec:partial_insurance} gives some detail on the challenges of partial insurance when $\underline{u}$ is unknown to the researcher. Some interesting implications of the model are discussed at the end.

\subsection{Researcher's uncertainty about the implementation}
\label{sec:impl_uncertain}

The researcher can have two types of uncertainty about the implementation that determines the distribution of the policy outcome, $X_m.$  Uncertainty about this distribution affects the researcher's willingness to offer a guarantee.  The first question is how well the implementation population matches the study population.  The second question is how closely the to-be-implemented policy matches the policy that was studied by the researcher.  Extending Section \ref{sec:illustration}'s illustration, both of these problems can be represented as $X_m \sim \mathrm{Binomial}(m, p q)$  for some $q \geq 0$ instead of $X_m \sim \mathrm{Binomial}(m, p)$.

Assessment of statistical results' external validity and their generalizability or transportability  to the implementation population \citep{degtiar_review_2023} resolves the first question.  The researcher must work with the implementer to estimate a lower bound for $q$ in the Binomial illustration.  A researcher who does not participate in this is probably insensitive to the policy outcome. 

The second question is a matter of moral hazard among implementers once their policy outcome is insured \citep{arrow_economics_1985}.  In some instances, the researcher would implement the policy on behalf of the implementer or closely supervise the implementer's actions in order to ensure the implementation matches the studied policy.  These can be expensive for the researcher. For simple policies, the researcher might audit the implementation after a policy failure and refuse to pay the guaranteed amount if the implementation does not match the study.  More complex policies and data with privacy restrictions are costly or impossible for the researcher to audit. 
Offering partial insurance of actual costs incurred (instead of projected) is a solution for these cases.  
If the implementer is not fully insured for their losses, then the implementer would want to maximize $q$ in order to maximize $\mathrm{E}[b(X_m)]$ under $X_m \sim \mathrm{Binomial}(m, p q).$

\subsection{Implementer's uncertainty about costs and benefits}
For simplicity's sake, the present model assumes costs, $c_m$ ($m = 1,\ldots,M$) and benefits, $b(x)$ ($x = 1,\ldots,M),$ of the policy are known to the implementer before implementation.  In reality, these costs and benefits are usually forecasts for which the actual values are not observed until after the policy is implemented.  \citet*{flyvbjerg_cost-benefit_2021} observe that these forecasts are often ``worse than worthless'' (p. 403) for large public infrastructure projects.  Reference class forecasting (RCF) remedies this by considering the distribution of realized costs and benefits in similar, already-implemented policies when making an implementation decision \citep*{flyvbjerg_curbing_2008,kahneman_intuitive_1979,lovallo_delusions_2003}.  RCF can also be combined with a statistical model of costs or benefits for the policy \citep{bordley_reference_2014}.  

Beyond recommending the use of RCF, \citet*{flyvbjerg_how_2005} suggest that ``forecasters and their organizations must share financial responsibility for covering benefit shortfalls (and cost overruns)'' (p. 143). \citet{flyvbjerg_cost-benefit_2021} give examples of forecasters being sued for wrong estimates.
As an alternative to suing, the present approach asks researchers to guarantee part of the benefit forecast. This can be extended to ask forecasters to guarantee their monetized benefit ($b(x)$) and cost ($c_m$) forecasts. Asking forecasters for a guarantee \emph{before} the implementation decision could be more efficient than suing after the fact because it would avoid unpredictable litigation and screen out less egregious cases that are likely to have a bad outcome, but are unlikely to win in court.

Cost and benefit forecasts cannot be guaranteed if they are unobservable after a policy is implemented.  Specifically, it may be impossible to guarantee a policy's opportunity cost relative to the status quo.  
This problem occurs in novel, dynamic situations like the first months of the COVID-19 pandemic when the status quo has been observed for only a short time. The implementer knows that forecasts of future status quo outcomes based on such a limited observation window are highly uncertain. However,  
once a new policy is implemented widely, there is no way to compare  forecasts of the status quo to what would have actually happened in the absence of the policy.

\subsection{Managing risk with multiple researchers}
\label{sec:multi_risk}

In the discussion so far, the implementer faces uncertain reliability of a single study as if it was the only estimate of a particular policy's effect.  When multiple studies estimate the effect of the same policy, the implementer can ask the researchers behind these studies to produce and collectively guarantee a single estimate.  The authors of highly unreliable estimates are unlikely to participate.  The remaining researchers can conduct a formal evidence synthesis, such as a meta analysis.  The resulting consensus estimate should be more reliable than any individual estimate. Furthermore, the participating researchers can share the risk of the policy outcome as in Section \ref{sec:researcher_decision}, so that each researcher faces less risk than if they had insured only their own study result. 

The same reasoning applies when an implementer is considering an existing consensus estimate. It is costly or impossible for an implementer to understand the quality of evidence synthesis or to uncover hidden biases.  A guarantee from the researchers who produce a consensus estimate can avoid the types of bias described in \cite*{kepp_panel_2024}, for example.

\subsection{Adverse selection against reliability}
\label{sec:researcher_adverse}
Section \ref{sec:implementer_decision} presents adverse selection as the implementer's failure to implement a policy that would be beneficial.  However, before an implementer sees any policy research, adverse selection can also work on the researcher side to prevent publication of beneficial results.
To illustrate this, suppose a researcher obtains the $(1-\alpha')\%$ confidence lower bound $L$ for the effect of a policy.  Suppose also that the researcher has in mind a probability distribution over $(c_1, \underline{u})$ for potential implementers of the policy.  Finally, suppose the researcher has an estimate of $\alpha,$ the false positive probability that implementers expect for this type of research.  The researcher faces a cost for publishing their results in terms of writing a manuscript, submitting it to journals, and responding to peer review.  If $\mathrm{Pr}(-c_1 \alpha < \underline{u})$ is large, the policy is not likely to be implemented. The value to the researcher of a never-implemented policy might not justify the cost to publish their results. If they do not bother to publish, implementers never have the opportunity to consider their results.

Researchers who are insensitive to the policy outcome might publish the same result with a smaller value of $\alpha'$ in order to increase the probability of implementation.  If this behavior is prevalent, implementers would respond to policy failures by lifting their expectations for $\alpha$ even further away from $\alpha'$, making future research even less likely to be implemented.  The reliability of published studies becomes increasingly overstated as this cycle repeats.
Thus, the total effect of adverse selection comes from a combination of decisions among implementer populations and researcher populations responding to each other over time.  

\subsection{Partial insurance as a signal}
\label{sec:partial_insurance}
A researcher's guarantee of the implementer's outcome is an extension of Spence's (\citeyear{spence_job_1973}) signaling model \citep[p.~114]{salanie2005economics}.  The guarantee is expensive for producers of unreliable results and inexpensive for reliable results.  This makes unreliable researchers less likely to offer a guarantee.  Therefore, the researcher's act of offering a performance guarantee sends information about reliability to the implementer.  It shows that the researcher is confident in both their estimate and the implementation.

The signaling model is especially relevant when the researcher is only willing to insure the implementer for tail losses exceeding $k.$  As mentioned, the implementer is willing to implement the policy if $k \geq \underline{u}.$  When $k < \underline{u},$ the implementer must make a decision using the value of $k$ as the signal.  In Section \ref{sec:ill_no_guarantee}, where there was no guarantee, the implementer considered their false positive probability, $\alpha,$ over the entire population of researchers.  Now the implementer can consider the false positive probability as a function of $k.$ Let $\alpha(k)$ be the expected false positive probability among the population of researchers who insure losses beyond $k.$   Since researchers with less reliable estimates pay a high price for these guarantees, the implementer can assume $\alpha(-c_m) = \alpha$ and $\alpha(k)$ is decreasing in $k.$  Using the worst-case decision rule from Section \ref{sec:ill_no_guarantee}, the implementer would implement the policy for $m$ such that $-\alpha(k) c_m \geq \underline{u}.$  Since $-\alpha c_m < -\alpha(k) c_m$, adverse selection is mitigated by transmitting the signal, $k.$  A similar argument applies to partial insurance of a proportion $s$ of the total loss.

The researcher's ability to send the implementer a reliability signal raises the question of how much insurance the researcher should offer relative to $\underline{u}.$  The possibility that the researcher might not know the implementer's value for $\underline{u}$ complicates the answer.  If the researcher asks the implementer for their value of $\underline{u},$ the implementer potentially benefits from responding  untruthfully. If the implementer responds with some $\underline{u}' > \underline{u}$, then the researcher might be willing to offer a larger amount of insurance in order to see their policy implemented.  This is not a problem if the implementer is a public agency with $\underline{u}$ defined by mandate. This is also not a serious problem for highly reliable research with a low false positive probability and a policy implementation that is easily audited. The situation where both the researcher and implementer can substantially benefit from dishonesty is more challenging and left for future discussion.

\subsection{Risk sharing motivates better peer review}
Section \ref{sec:researcher_risk_mgmt} discussed two ways that researchers could transfer their risk to a third party in order to bring that risk to a bearable level at which they could guarantee the policy outcome.  Whichever the transfer method, the third party would need to understand the distribution of $Y_m$ and $L$ before agreeing to the transfer. This requires the third party to have domain knowledge and statistical expertise along with access to the study data and analysis code.  The risk transfer implies a deeper and more rigorous peer review than what is often required of journal article referees.

\subsection{Institutional risk management}
Research institutions such as universities can provide a more efficient system for the risk transfers of Sections \ref{sec:researcher_risk_mgmt} than individual researchers transacting between each other. First, the volume of research produced by institutions amounts to a more diversified risk portfolio than that of a small number of researchers within the same field sharing risks from a small number of policy outcomes.  Second, dedicated personnel within the institution can do the administrative work of risk management instead of each individual researcher working on their own.  

Research institutions can also mitigate adverse selection more effectively than researchers alone.  First, the institutional guarantee of a policy outcome signals the institution's confidence in their researcher's output.  This could make implementers more willing to implement.  Second, although discussion has been omitted so far, the implementer must consider counter-party risk.  An individual researcher might not have sufficient wealth to cover their guarantee after a policy failure. Institutional backing makes this less risky for implementers.  This is especially true for policy outcomes realized over a longer time horizon than the researcher's remaining career.

\subsection{Application beyond policy implementation}
While this paper considered risk management in policy implementation, the ideas extend more broadly to any decision that depends on statistical results.  As an example, \citet{osherovich_hedging_2011} discusses ``academic risk'' hedging in the context of translational biomedical research.  Startup companies rely on academic studies to identify promising candidates for commercial development. The companies embark on costly research with the expectation that results from roughly half of studies will fail to replicate.  These companies could manage risk by asking academic researchers to insure some of the company's cost to validate the research results. 
\citet{rosenblatt_incentive_2016} proposes that research institutions offer a ``money-back guarantee.''
By considering results from only the researchers and institutions who are willing to participate in such a guarantee, the replication rate should improve, and, among the validation studies that fail, the companies hedge their costs.

\subsection{Bottom-up versus top-down change}
The fundamental problem discussed in this paper is the implementer's perception of the reliability distribution among researchers.  The performance guarantee allows individual researchers to signal unilaterally that they are part of a more reliable distribution.  The success of this approach does not require widespread participation among the researcher population.  However, once a few researchers participate, implementers may come to demand that all researchers guarantee outcomes as a requirement for having their policies implemented.  This could push other researchers toward making greater investments into reliability.  In this way, the initial actions of a few researchers would improve the reliability distribution from the bottom up. This might be a more efficient way to change the distribution than top-down approaches that try to shift a critical mass of researchers simultaneously toward greater reliability.  Top-down examples include attempts to change systems of research training, research funding, publication, or academic employment.

%\bigskip
%\begin{center}
%{\large\bf SUPPLEMENTARY MATERIAL}
%\end{center}
%
%\begin{description}
%
%
%\item[R program for Example 2 false positive probability:] R program containing code to produce the plots in Figure \ref{fig:ex2fpr}. (.R file)
%
%
%\end{description}

%\spacingset{1}

\bibliographystyle{apalike}

\bibliography{adverse_selection_statistical_practice}
\end{document}